\documentstyle[12pt]{article}
\begin{document}
\renewcommand{\thefootnote}{\fnsymbol{footnote}}
\newcommand{\bfx}{{\bf x}}
\newcommand{\bfy}{{\bf y}}
\newcommand{\bfr}{{\bf r}}
\newcommand{\bfk}{{\bf k}}
\newcommand{\bkp}{{\bf k'}}
\newcommand{\beq}{\begin{equation}}
\newcommand{\eeq}{\end{equation}}
\newcommand{\beqn}{\begin{eqnarray}}
\newcommand{\eeqn}{\end{eqnarray}}
\newcommand{\beqnn}{\begin{eqnarray*}}
\newcommand{\eeqnn}{\end{eqnarray*}}
\rightline{\large\baselineskip16pt\rm\vbox to20pt{
           \hbox{OCHA-PP-73}
           \hbox{January 1996}}}
\begin{center}
{\large\bf Electroweak baryogenesis \\ with the vector-like quark model}
\footnote{To appear in the Proceedings of the workshop on 'Electroweak 
Baryogenesis' held at Chikuho on November 25(1995). This talk is based on the
collaboration with A. Sugamoto and A. Yamaguchi (Ochanomizu University).}
\end{center}
\begin{center}
{\bf{Tomoko Uesugi} \\
\sl{Department of Physics, Ochanomizu University, Tokyo 112, Japan}}
\end{center}
%
\begin{abstract}
  We study on the electroweak baryogenesis problem in the vector-like 
quark model. This model can add to the minimal standard model the extra 
CP-violation source which breaks CP spontaneously. Using charge 
transport mechanism suggested by Nelson et al, we can get 
the net hypercharge flux : $f_{Y} \sim 10^{-5}$. This result
could nicely explain the present baryon to entropy 
ratio : $\rho_{B}/s \sim 10^{-9}$. 

\end{abstract}  
%
\section{Introduction}
Recently, Nelson, Kaplan and Cohen suggested the electroweak 
baryogenesis scenario
called the charge transport mechanism \cite{nkc}.
Using this mechanism, we study on the baryogenesis problem 
in the vector-like quark model. 
This model was first introduced by Bento, Branco and Parada 
\cite{bbp} as one of the most simplest spontaneous CP-violating 
models.
 The vector-like quark model is attractive to the electroweak 
baryogenesis because it can 
not only add to the standard model an extra CP violation 
source, but also make the first order phase transition of the
electroweak theory much stronger.

\section{The vector-like quark model}
  
We consider the extension of the standard model by adding extra SU(2) 
singlet quarks $U$ as the fourth generation as well as an extra 
singlet Higgs $S$.

 The field content of this model is 
$( u \ d )^{i}_{L}$, $u^{\alpha}_{R}$, $d^{i}_{R}$, $U_{L}$, $\phi$, 
$S$, $( i=1,2,3, \  \alpha=1,2,3,4)$, 
 where $i$ and $\alpha$ are the generation indices, $\phi$ denotes 
the standard Higgs doublet. We assume that the additional singlet 
Higgs $S$ is a complex scalar and all the new fields introduced 
are odd under a $Z_{2}$ symmetry.

Under these conditions, the $SU(2)\times U(1)\times Z_{2}$ invariant 
Yukawa couplings are
  ${\cal L}_{Y}=-\sqrt{2}(\bar{u} \bar{d})^{i}_{L}(h_{ij}\tilde{\phi}d^{j}_{R}
           +f_{ij}\phi u^{j}_{R})-\mu\bar{U}U
           -\sqrt{2}(f_{i4}S+f^{'}_{i4}S^{*})\bar{U_{L}}u^{i}_{R}+h.c. $
 
 During the phase transition, $\phi$ and $S$ are able to have vacuum 
expectation values as $\langle \phi \rangle =\frac{1}{\sqrt{2}} 
\pmatrix{ 0 \cr v \cr} $ , 
$\langle S \rangle = \frac{V}{\sqrt{2}} e^{i \alpha}$. 
We consider the most general renomalizable $SU(2)\times U(1)\times Z_{2}$
symmetric scalar potential. Which can be written as

$ V_{\phi,S}=\frac{\rho^{2}}{2}\phi^{+}\phi+\frac{\lambda}{4}
       (\phi^{+}\phi)^{2}+S^{*}S(a_{1}+b_{1}S^{*}S) 
       +(S^{2}+S^{*2})(a_{2}+b_{2}S^{*}S)$

$ \ \ \ \ \ \  +b_{3}(S^{4}+S^{*4}) 
       +\phi^{+}\phi[c_{1}(S^{2}+S^{*2})+c_{2}S^{*}S ]$.

In order to violate CP spontaneously, this potential must have the 
minimum value with the non-zero complex phase.
The $\alpha$-dependent part of the potential has the minimum value at
$\alpha=\frac{1}{2}\cos^{-1}\left[-\left(a_{2}+\frac{b_{2}V^{2}}{2}
                +\frac{c_{1}v^{2}}{2} \right)/(b_{3}V^{2}) \right]$,
where $v$ is space-dependent inside the bubble wall, so that $\alpha$ 
must also have the space dependence inside the wall.
The shape of the wall $v(z)$ can be obtained from the equation 
of motion of $\phi$,
 but here, we assume a wall profile simply as 
$v(z)=v_{0}\left(1+\tanh(z/\delta_{w})\right)/2$ .
We consider only the z-dependent solution, where the z axis is normal 
to the bubble 
wall, since we can get general solutions by boosting the wall profiles
in the x-y plane.
$\delta_{w}$ is the parameter which determines the thickness of the 
wall.

\section{Computation of the hypercharge flux} 
For simplicity, we treat only top and vector-like quarks to compute 
net hypercharge flux. What we have to do is only to solve the Dirac 
equation.  
The two-generations' up quark mass matrix is
\beqnn
    M=\pmatrix{fv & 0 \cr FVe^{i\alpha}+F^{'}Ve^{-i\alpha} & \mu \cr}
\eeqnn

Using this mass matrix, we can write the Dirac equation in the chiral bases.
For the stationary state($ \psi(t,z) = \psi(z) e^{iEt} $), the Dirac equation
can be written by
\beqnn
   i\frac{\partial}{\partial z}\pmatrix{\psi_{1R}(z) \cr \psi_{3L}(z) \cr}
        = -\pmatrix{E & M^{+} \cr M & -E \cr }\pmatrix{\psi_{1R}(z) 
                                 \cr \psi_{3L}(z) \cr} \nonumber  \\
   i\frac{\partial}{\partial z}\pmatrix{\psi_{4L}(z) \cr \psi_{2R}(z) \cr}
        = -\pmatrix{E & M^{+} \cr M & -E \cr }\pmatrix{\psi_{4L}(z)
                                 \cr \psi_{2R}(z) \cr} 
\eeqnn   
where $\psi_{L(R)}(z)$ are two component spinors, corresponding to 
the top and the vector-like quarks.
We can solve this Dirac equation numerically as a simple system of ordinary  
differential equations with respect to z. Using the reflection 
coefficient from right-handed to left-handed, $R(E)$, and from 
left-handed to right-handed, 
$\bar{R}(E)$, the hypercharge flux can be written by the integration over
energy of the incident particles.
For simplicity, we neglect the final distribution term, and 
the integral over transverse momentum $k_{T}$ can be done analytically.
Using this approximation for the hypercharge flux, and setting 
$\mu=600 [GeV]$, $m_{t}=174 [GeV]$, $m_{H}=100 [GeV]$ and 
$v_{w}=0.5 \times c$, 
we can get the final result for the computation of the hypercharge flux, 
 $f_{Y}\sim 10^{-5}$ $[T^3]$. This value can be obtained by taking 
reasonable values for the various couplings in the scalar potential and 
for the mass of the singlet Higgs.

\section{Summary}
Using the charge transport mechanism, we could nicely explain present
baryon to entropy ratio with the vector-like quark model.
This model is able to add new source of CP violation as a spontaneous CP 
violation. In this case, the CP violating phase becomes space-dependent,
and the hypercharge flux can be produced in the bubble wall.
The produced hypercharge is finally converted to baryon number excess 
through the sphaleron transition as usual.

For details, see the forthcoming paper \cite{tys}.


\begin{thebibliography}{22}
\bibitem{nkc}
A. E. Nelson, D. B. Kaplan and A. G. Cohen, Nucl. Phys.
{\bf B373}(1992),453;
A. G. Cohen, D. B. Kaplan and A. E. Nelson,
Ann. Rev. and Part. Sci., {\bf 43}(1993),27.
\bibitem{bbp}
L. Bento, G. Branco and P. A. Parada, Phys. Lett. {\bf B267}(1991),95. 
\bibitem{tys}
T. Uesugi, A. Yamaguchi, A. Sugamoto \\
 {\it Electroweak baryogenesis 
with the vector-like quark model} (in preparation).


\end{thebibliography}
\end{document}